\documentstyle[12pt]{article}
\title{Phase diagram in 2D Fr\" ohlich model of metal at arbitrary carrier 
density:  pseudogap {\it versus} doping}
\author{{\sl V.M.~Loktev, S.G.~Sharapov and 
V.M.~Turkowski$^{\dagger}$}\\ 
{\sl  Bogolyubov Institute for Theoretical Physics,}\\ 
{\sl Metrologichna str. 14-b, Kyiv, 252143 Ukraine}\\
{\sl $^{\dagger}$ Kyiv Shevchenko University}\\ 
{\sl Acad. Glushkova prosp. 6, Kyiv, 252127 Ukraine}} 
\date{}
\begin{document}
\maketitle


\begin{abstract}
It is shown that in 2D system of the fermions with simplest indirect 
boson-induced attraction (through the Einstein phonon exchange as an 
example) along with the normal and superconducting phases there arises 
a new (called "abnormal normal" or pseudogap) one where the absolute 
value of the order parameter is finite but its phase is a random 
quantity.  It is important that this new phase really exists at low 
carriers density only, i.e. it shrinks when doping increases.  The 
relevance of the results obtained with dependence of pseudogap on 
doping in high-temperature superconductors is speculated.  
\end{abstract}



{\sl* Corresponding author: V.M.~Loktev \\
Bogolyubov Institute for Theoretical Physics,\\ 
Metrologichna str. 14-b, Kyiv, 252143 Ukraine\\
E-mail: vloktev@gluk.apc.org}

\eject

\section{Introduction}

An adequate description of the physical properties of 
high-temperature superconductors (HTSCs) is one of the most important 
problems of the modern solid state physics (see, for example, review 
\cite{Lok1}).  And among the most debatable questions on 
HTSCs is the question about the so called pseudogap (or spin gap) 
which is experimentally observed in normal state samples with lowered 
carrier density $n_f$. The matter is that HTSCs represent systems 
with relatively easy changeable value of $n_f$ what is interesting and 
important both from experimental and from theoretical points of view. 
In particular, the underdoped samples reveal a lot of strange 
anomalies concerning their spectral, magnetic, thermodynamic and 
other observables (among them there is the pseudogap which has 
been seen in ARPES experiments \cite{Mar1,Din1,Loes1} and now widely 
discussed \cite{Lev1,Bat1,Ong1}). Moreover, even the definition 
itself "underdoped" ("overdoped") HTSC sample 
is related to the presence (absence) of such anomalies.

On the other hand, the possibility of changing $n_f$ value is crucial
for the theory and puts the general problem of the crossover from 
composite boson superfluidity (underdoped regime, or small $n_f$)
to Cooper pairing when $n_f$ increases (overdoped regime).
This crossover was already studied for the systems of different
dimensionalities:  
3D \cite{Ran1} and quasi-2D 
\cite{Gor1}. As for 2D case, it has been considered for 
temperature $T=0$ only \cite{Ran1,Gor2} what is conditioned 
by the Hohenberg-Mermin-Wagner theorem which forbids any 
long-range order in such systems at finite $T$ because of long-wave 
fluctuations of the charged order parameter.

Formation of inhomogeneous condensate, or 
the Berezinskii-Kosterlitz-Thouless (BKT) phase, is possible, but 
its consistent study is connected with some difficulties. 
Nevertheless, this case was already explored in 2+1
relativistic field model \cite{Mac1} where, as it is 
known, concentration effects are not justifiable.  They 
are known for some nonrelativistic models (see, for example, 
\cite{Drec1}), but the BKT equation in this paper was obtained 
without taking into account the existence of neutral order 
parameter $\rho$ ($\not= 0$). Its introduction and use, as it was 
shown in \cite{Gus1}, is very important and leads to formation of a 
new phase with $\rho\not= 0$ which separates normal phase and 
another one, which is also normal, since superconducting 
properties are absent there as well. 

Below an attempt is made to study the above mentioned crossover and 
the new phase formation possibility. In contrast to 
\cite{Gus1}, where this question was studied for a 2D four-fermion 
(4F) model, we shall consider a more realistic Fr\"ohlich model.  It 
will be shown that the unknown phase in this case not just exists, 
but appears mainly at low carrier density what is 
essentially different from the 4F-case.  Such a result 
promises that the models with an indirect attractive femion-fermion 
interaction may be suitable to account for the unusual features in 
the normal phase behaviour of HTSCs.

\section{Model and main equations}

Let us choose the Fr\"ohlich model Hamiltonian density in the 
standard form:  
\begin{equation} H(x) =-\psi_{\sigma}^{\dagger}(x) 
      \left(\frac{\nabla^2}{2m}-\mu\right)\psi_{\sigma}(x) + 
      g\varphi (x)\psi_{\sigma}^{\dagger}(x)\psi_{\sigma}(x) + 
H_{ph}(\varphi (x)), (x=\mbox{\bf r},t), \label{1} 
\end{equation} 
where $\psi_{\sigma}(x)$ is a
fermion field; $m$ and $\sigma =\uparrow ,\downarrow $ are an 
effective mass and a spin of the fermions, $\mu$ is their 
chemical potential which fixes $n_{f}$; $\varphi (x)$ is a phonon 
field operator and $g$ is a fermion-phonon coupling constant.  We use 
Pauli matrices $I,\tau_x,\tau_y,\tau_z$ and put $\hbar=k_B=1$.

In (\ref{1}) $H_{ph}$ is the Hamiltonian of free phonons, which 
results in the simplest expression for the phonon propagator 
(in the Matsubara temperature formalism)\cite{Abr1}
\begin{equation} 
D(i\Omega_n ) 
=-\frac{\omega_0^2}{\Omega_n^2+\omega_0^2} , 
\qquad    \Omega_n=2n\pi T \label{2} 
\end{equation} 
with $\omega_0$ as 
Einstein phonon frequency and $n$ as an integer. In principle, 
(\ref{2}) corresponds to a propagator of any massive bosonic 
excitations by which fermions can exchange\footnote{In spite of its 
simplicity such a model is rather close to HTSCs with their developed 
quadrupolar mode spectra - optical phonons and $dd$-excitons 
\cite{Lok1}.  Exchange of these excitations leads to short-range 
interaction between fermions.  It, however, does not include spin 
excitations (almost massless in HTSCs) which can provide the 
long-range inter-carrier interaction what, of course, needs a 
separate study.}.

With the purpose to calculate the phase diagram it is necessary to
find the thermodynamic potential of the system. 
In the case of the 4F model it can be 
obtained by making use of the well-developed Hubbard-Stratonovich 
method in which the statistical sum $Z$ can be represented by a path 
integral over the fermionic 
$\psi_{\sigma}(x)$ and the 
complex Hubbard-Stratonovich $\phi (x) 
=V<\psi_{\uparrow}^{\dagger}\psi_{\downarrow}^{\dagger}>$ 
fields ($V$ is the 4F-interaction constant). 

As it was shown earlier \cite{Tau1,Gus1}, in 2D case it is 
convenient to pass from $\phi$ and $\phi^{*}$ fields to new 
variables, namely: the absolute value $\rho$ and the phase 
$\theta$, where  $\phi (x)=\rho (x) \exp [-2i\theta (x)]$, 
and to perform simultaneously the spinor transformation 
\begin{equation} 
\psi_{\sigma} (x)= \chi_{\sigma} (x) 
\exp [i\theta (x)], \label{3} 
\end{equation} 
where $\chi_{\sigma}(x)$ corresponds 
to a neutral fermi-particle field.  Such a substitution allows to 
represent $Z$ in the following form 
\begin{equation} 
Z=\int\rho {\cal D}\rho {\cal D}\theta 
\exp\left[-\beta\int\Omega(\rho (x) ,\theta (x))dx\right] , \ \  
(\beta =1/T) \label{4} 
\end{equation} 
where $\Omega$ is the 
effective quantum thermodynamic potential in terms of $\rho (x)$ and 
$\theta (x)$ variables.  This representation proves to be very useful 
and leads to  (at $\rho (x)=const$ and expansion in $\nabla \theta 
(x)$ up to $(\nabla \theta (x))^{2}$) the effective Hamiltonian which 
is similar to that of the XY-model.  As  a result, the equation for 
$T_{BKT}$ in the latter model can be directly used for the case under 
consideration.  The solution of whole the set of the self-consistent 
equations for $\rho$, $\mu$ and $T_{BKT}$ was given in \cite{Gus1}.

This approach is however inapplicable in the more complex case of 
indirect interaction model for which the local Hubbard-Stratonovich 
fields can not be introduced. At the same time, by making use 
of the Cornwall-Jackiw-Tomboulis formalism \cite{Cor1} 
it is possible to obtain classical thermodynamic potential which 
depends on bilocal  
$<\psi_{\downarrow} (x)\psi_{\uparrow} (y)>$ and 
$<\psi_{\uparrow}^{\dagger}(x)\psi_{\downarrow}^{\dagger}(y)>$ 
fields (or, more exactly, on the full fermion Green function (see 
below)) though, strictly speaking, this potential does not  satisfy 
Eq.  (\ref{4}) exactly. The corresponding effective action 
(equivalent to $\Omega$ in (\ref{4})) in two-loop approximation can 
be calculated and (in the Nambu representation \cite{Sch1}) takes the 
form 
\begin{equation} 
\beta\Omega [{\cal G}]=- \mbox{Tr} \left(\mbox{Ln} 
{\cal G}^{-1} {\cal G}_{0}+{\cal G}{\cal G}_{0}^{-1}-1\right)+ 
\frac{g^2}{2}\mbox{Tr}{\cal G}\tau_zD\tau_z{\cal G}, \label{5} 
\end{equation} 
where Tr includes integration over 2D
space $\mbox{\bf r}$ and imaginary time $0\leq\tau\leq\beta$ 
as well as the standard trace operation; 
\begin{eqnarray} 
{\cal G}^{-1} 
& = & -I\partial_{\tau}+\tau_z\left(\frac{\nabla^2}{2m}+\mu\right) 
+\tau_{+}\phi^{*}+\tau_{-}\phi;
\label{Green} \\
{\cal G}_0^{-1} & = & {\cal G}^{-1}(\rho =0) \nonumber
\end{eqnarray}
are the full and free fermion Green functions, 
and $D$ was defined in (\ref{2}). We also use the normalization 
condition ${\cal G}_0={\cal G}_0(\mu =0)$ under the symbol Ln
in (\ref{5}).

The stationary condition $\delta \Omega({\cal G}) / \delta {\cal G} 
= 0$ results in the well-known Schwinger-Dyson equation
\footnote{In fact it coincides with the Eliashberg equation 
\cite{Sch1} when the renormalization factor is $Z( i \omega_{n}) = 1$.}
for the inverse full fermion Green function 
\begin{equation} 
{\cal G}^{-1}= {\cal G}_{0}^{-1}-g^2\tau_z {\cal G}\tau_zD. \label{7} 
\end{equation} 
Substituting (\ref{7}) into (\ref{5}) one can obtain a simpler 
expression for $\Omega ({\cal G})$ where the phonon Green
function (\ref{2}) is already omitted
\begin{equation}
\beta\Omega ({\cal G}) =-\mbox{Tr}\left[\mbox{Ln}{\cal G}^{-1}{\cal 
G}_{0}+\frac{1}{2}\left({\cal G}{\cal G}_{0}^{-1}-1\right)\right]. 
\label{8} 
\end{equation}  

To find the potential $\Omega $ as a function of phase 
gradient $\nabla\theta (x) $ and absolute value $\rho $ it is 
necessary to make the transformation (\ref{3}).  Then one  
obtains for (\ref{Green}):
\begin{eqnarray} 
{\cal G}^{-1} & = & -I\partial_{\tau}+\tau_z\left(\frac{\nabla^2}{2m}+\mu\right)+ 
\tau_x\rho-  \nonumber \\
&& \tau_z\left(\partial_{\tau}\theta+ 
\frac{\nabla\theta^2}{2m}\right)+iI\left(\frac{\nabla^2\theta}{2m}+
\frac{\nabla\theta\nabla}{m}\right)\equiv
G^{-1}(\rho ) -\Sigma (\partial\theta );   \label{9} \\ 
 {\cal G}_0^{-1} & \equiv  & G_{0}^{-1}-\Sigma 
 (\partial\theta).  \nonumber
 \end{eqnarray} 

Since the low energy dynamics is mainly determined by the phase 
fluctuations and corresponds to the region where spatially  
homogeneous order parameter $\rho\not= 0$ it is 
sufficient to be restricted the expansion in terms of $\Omega$ in 
$\nabla\theta (x)$ only; for instance, the Green function (\ref{9}) 
${\cal G}=G+G\sum_{n=0}^{\infty} (G\Sigma)^n$.  Then the 
desirable effective potential (\ref{8}) can be divided it two parts: 
$\Omega =\Omega_{pot}(\rho ) + \Omega_{kin}(\rho ,\nabla\theta )$, 
where in $(\nabla \theta)^{2}$ approximation
\begin{eqnarray}
\beta\Omega_{kin}(\rho ,\nabla\theta 
) & = &\mbox{Tr}\left[G\Sigma+\frac{1}{2}G\Sigma G\Sigma-G_0\Sigma-
\frac{1}{2}G_0\Sigma G_0\Sigma+\right.
                                         \nonumber  \\
& & \left.\frac{\rho}{2}\tau_{x} G 
( G \Sigma + G \Sigma G \Sigma) \right].\label{10} 
\end{eqnarray} 
Assuming then that $\rho ( i \omega_{n}) = const$ one obtains from 
(\ref{10}) after somewhat tedious but otherwise straightforward 
calculation 
\begin{equation} 
\Omega_{kin}=\frac{T}{2}\int_{0}^{\beta}d\tau\int d^2\mbox{\bf r} 
J(\mu, T, \rho (\mu ,T))(\nabla\theta)^2, \label{11} 
\end{equation} 
where 
\begin{eqnarray} 
J(\mu ,T,\rho (\mu ,T))=\frac{1}{2\pi}(\sqrt{\mu^2+\rho^2}+\mu +
2T\ln\left[1+\exp\left(-\frac{\sqrt{\mu^2+\rho^2}}{T}\right)\right]-
                                               \nonumber \\
\frac{T}{\pi}\left[1- \frac{\rho^2}{4T^2}
\frac{\partial}{\partial (\rho^2/4T^2)}\right]\int_{-\mu 
/2T}^{\infty}dx\frac{x+(\mu /2T)} 
{\cosh^2\sqrt{x^2+\rho^2/4T^2}}. \label{12} 
\end{eqnarray}
Note that in comparison with the 4F model \cite{Gus1}
the last expression contains one more term with the derivative.

The equation for the temperature $T_{BKT}$ of the BKT transition can 
be written after direct comparison of $\Omega_{kin}$ (\ref{11}) with 
the Hamiltonian of the XY-model which has the same form 
\cite{Izum1}; hence 
\begin{equation} \frac{\pi}{2}J(\mu ,T_{BKT},\rho 
(\mu ,T_{BKT})) = T_{BKT}.  \label{13} 
\end{equation} 
To complete a 
set of selfconsitent equations which allow us to trace the dependence 
of $T_{BKT}$ on $n_{f}$ the equations for $\rho$ and $\mu$ have 
also to be given.
      
      In particular the equation for $\rho(i \omega_{n})$
is nothing else but $(\ref{7})$ with $\nabla \theta = 0$,
i.e. the Green function $G$ of the neutral fermions replaces 
$\cal G$, so that $(\ref{7})$ in frequency-momentum represantation 
takes the form
\begin{equation}
\rho(i \omega_{n}) = 
g^2T\sum_{m=-\infty}^{\infty}\int\frac{d^2\mbox{\bf 
k}}{(2\pi)^2}\frac{\rho (i\omega_m)}{\omega_m^2+\varepsilon^2 
(\mbox{\bf k})+\rho^2 
(i\omega_m)}\frac{\omega_0^2}{(\omega_m-\omega_n)^2+\omega_0^2},
\label{14} 
\end{equation} 
where $\omega_n=(2n+1)\pi T$ is the Matsubara fermionic 
frequency \cite{Sch1}. 

Analytical study of this equation as well as obtaining both
(\ref{12}) and the number equation is possible only if one supposes 
that $\rho(i \omega_{n}) = const$
\footnote{For the case $T = 0$ this assumption was checked in 
\cite{Lok2}.}.

 Finally, the number equation which follows from the condition 
 $v^{-1}\partial\Omega[{\cal G}] /\partial\mu =-n_{f}$ (where 
 $v$ ia a volume of the system) and is crucial for crossover 
 description has to be added to (\ref{13}) and (\ref{14}) for 
 self-consistency; so one comes to 
\begin{equation} \sqrt{\mu^2+\rho^2}+\mu 
+2T\ln\left[1+\exp 
\left(-\frac{\sqrt{\mu^2+\rho^2}}{T}\right)\right]= 2\epsilon_F,
\label{15} 
\end{equation} 
where $\epsilon_F=\pi n_f/m$ is the Fermi energy of 2D fermions with 
the simplest quadratic dispersion law.  Thus, in the case under 
consideration all unknown quantities ($\rho ,\mu$ and $T_{BKT}$) are 
functions of $n_f$. 

\section{Analysis of the solutions}

It is natural to guess that in contrast to the accepted 
XY-model in the superconducting model there exists one more critical 
temperature, also dependent on $n_f$, where the complete order 
parameter vanishes. This critical temperature, $T_{\rho}$, can be 
found from (\ref{14}) and (\ref{15}) by putting $\rho =0$ (what in 
accordance with these equations derivation corresponds to the 
mean-field approximation). As a result, with temperature decreasing 
2D metal passes from normal phase ($T>T_{\rho }$) to another one 
where averaged homogeneous (charged) order \\
parameter  $<\phi (x)>=0$, or, what is the same, superconductivity is absent, 
but chargeless order parameter $\rho\not= 0$.  It is very important that 
the pseudogap is formed just in the temperature region 
$T_{BKT}<T<T_{\rho }$, because, as it follows from above formulas 
(see, for instance, (\ref{14}), (\ref{15})), $\rho$, which depends on 
$n_f$ and $T$, enters all spectral characteristics of 2D metal in the 
same way as the superconducting gap $\Delta (T)$ enters into 
corresponding expressions for ordinary superconductors.  It explains 
why this, new phase can be called the "abnormal normal" phase or, 
better, pseudogap one.  The density of states near $\epsilon_F$ in 
the pseudogap is definitely less than in the normal phase, but is not 
equal zero as in superconducting one.  

The phase diagram of the 
system is presented in fig.1.  Both curves for $T_{\rho }$ and 
$T_{BKT}$ as functions of $n_f$ can be obtained by numerical 
calculation.  As for their asymptotic behaviour can be established 
analitically.

Indeed, it is not difficult to be convinced that the asymptotics for 
$T_{\rho }(n_f)$ and $T_{BKT}(n_f)$ have the forms:

a) when ratio $\epsilon_F/\omega_0\ll 1$ (very low fermion
density, or local pair case) the first of them satisfies the 
equation $T_{\rho }\ln (T_{\rho }/\epsilon_F)= \omega_0\exp 
(-2/\lambda)$ which immediately results in $\partial T_{\rho}(n_f)/
\partial n_f |_{n_f\rightarrow 0}\rightarrow\infty$.  
At the same time the temperature $T_{BKT}$ here has has another 
carrier density dependence: $T_{BKT}=\epsilon_F/2$ what in its turn 
means that it is equal to the number of composite bosons. It means 
that in this density region $T_{\rho}/T_{BKT}\gg 1$.

b) in the case $\epsilon_F/\omega_0\gg 1$ (very 
large fermion density, or Cooper pair case) one comes to 
standard BCS value:  $T_{\rho}=(2\gamma\omega_0/\pi) \exp 
(-1/\lambda)\equiv T_{BCS}^{MF}=(2\gamma 
/\pi)\Delta_{BCS}$ ($\Delta_{BCS}$ is the one-particle 
BCS gap at $T=0$). In other words, the temperature 
$T_{\rho}$ becomes equal to its BCS value 
\footnote{Being equal (in 
mean field approximation only) these temperatures ($T_{\rho}$ and 
$T_{BCS}^{MF}$) are in fact different: if $T_{BCS}^{MF}$ immediately 
falls down to zero as fluctuations are taken into account, $T_{\rho}$ 
does not and is renormalized only.}.  
The $T_{BKT}$ asymptotics in this 
case is not so evident and requires more detailed consideration. 

Firstly, it is naturally to 
suppose that for large $n_f$ values $T_{BKT}\rightarrow T_{\rho}$. 
Then it is necessary to check 
the dependence of $\rho $ on $T$ as $T\rightarrow T_{\rho}$. 
For that the equation (\ref{14}) can be transformed to:  
$$
\frac{1}{\lambda} 
=\int_{0}^{\infty}dx\left(\frac{\tanh\sqrt{x^2+\rho^2/4T^2}} 
{\sqrt{x^2+\rho^2/4T^2}}-\frac{\tanh\sqrt{x^2+\rho^2/4T^2}-
\tanh (\omega_0/2T)}{2(\sqrt{x^2+\rho^2/4T^2}-\omega_0/2T)}-\right. 
$$
\begin{equation}                                                     
\left.\frac{\tanh\sqrt{x^2+\rho^2/4T^2}+\tanh 
(\omega_0/2T)} {2(\sqrt{x^2+\rho^2/4T^2}+\omega_0/2T)}\right) 
\label{16} 
\end{equation} 
(where it was used that in this concentration region 
the ratio $\mu /2T_{\rho}\simeq\epsilon_F/2T_{\rho} \gg 1$).

Because  usually $\omega_0/2T_{\rho}\gg 1$, only very small $x$ 
give the main contribution to the integral (\ref{16}) (it is seen 
from the limit $\rho /2T_{\rho} \rightarrow 0$ when 
 $\epsilon_F/\omega_0\rightarrow\infty$). Therefore the latter 
 expression takes the approximate form:  
\begin{equation} 
\frac{1}{\lambda}=\int_{0}^{\infty}dx\left(\frac{\tanh\sqrt{x^2+\rho^2/4T^2}}
{\sqrt{x^2+\rho^2/4T^2}}-\frac{1}{x+\omega_0/2T}\right).\label{17} 
\end{equation}
 On the other hand, the condition $\rho =0$ in (\ref{17}) leads to
the equation
\begin{equation} 
\frac{1}{\lambda}=\int_{0}^{\infty}dx\left(\frac{\tanh x}{x}-
\frac{1}{x+\omega_0/2T_{\rho}}\right).\label{18}
\end{equation}
for $T_{\rho}$.
From (\ref{17}) and (\ref{18}) it directly follows that 
$$
\int_{0}^{\infty}dx\left(\frac{\tanh 
x}{x}-\frac{\tanh\sqrt{x^2+\rho^2/4T^2}} 
{\sqrt{x^2+\rho^2/4T^2}}\right)=\ln\frac{T_{\rho}}{T}.
$$
Then using the approximation
\[
\frac{\tanh \sqrt{x^2+\rho^2/4T^2}}{\sqrt{x^2+\rho^2/4T^2}}\simeq 
\left\{ 
\begin{array}{ll}
1-3^{-1}\left[x^2 +\rho^2/4T^2\right], & x \leq 1 \\ 
x^{-1} -\rho^2/8T^2x^3, & x>1
\end{array}
\right.  
\]
one easily comes to expression needed: 
\begin{equation} 
\rho\simeq 2.62T_{\rho}\sqrt{T_{\rho}/T-1}.\label{19} 
\end{equation} 
Remind that the generally accepted 3D result is $\Delta_{BCS}(T) 
=3.06T_{BCS}^{MF}\sqrt{T_{BCS}^{MF}/T-1}$ \cite{Abr1} and this small
difference can be explained by the above approximation 
what, however, is suitable for the following below 
qualitative discussion (see next Section).  

The last dependence has to be substituted in equation (\ref{13}).
And because of $\mu 
/2T_{BKT}\simeq\epsilon_F/2T_{BKT}\gg 1$, and $\rho 
(T_{BKT})/2T_{BKT}\ll 1$ when $T_{BKT}\rightarrow 
T_{\rho}$ this equation can be written as 
\begin{eqnarray} 
\frac{\epsilon_{F}}{4T_{BKT}}\left[1-\frac{\rho^2}{4T_{BKT}^2}
\frac{\partial}{\partial (\rho^2/4T_{BKT}^2)}\right]
\int_{0}^{\infty}dx\left(\frac{1}{\cosh^{2}x}-\right.
                                                   \nonumber \\
\left.\frac{1}{\cosh^{2}\sqrt{x^2+\rho^2/4T_{BKT}^2}}\right)=1.
\label{20}
\end{eqnarray}
At last, using expansion in $\rho /2T_{BKT}$ in integral 
(\ref{20}), the latter can be transformed to 
\begin{equation} 
\frac{a\epsilon_{F}}{8T_{BKT}}\left(\frac{\rho}{2T_{BKT}}\right)^4=1,
\label{21} 
\end{equation} 
where the numerical constant
$$ a=\int_{0}^{\infty}dx\frac{\tanh^2x-x^{-1}\tanh x+1}
{2x^2\cosh x}\simeq 1.98.  
$$ 
Combining now (\ref{19}) and (\ref{21}) one comes to the final 
relation between $T_{\rho}$ and $T_{BKT}$ for large carrier 
density:
\begin{equation} 
T_{BKT}\simeq 
T_{\rho}(1-1.17\sqrt{T_{\rho}/\epsilon_{F}}),\label{22} 
\end{equation} 
i.e. $T_{BKT}$ as a function of $n_f$ approaches $T_{\rho}$
(or $T_{BCS}^{MF}$).

\section{Discussion}

Thus, in the considered model of 2D metal with indirect interection 
of carriers, along with the normal phase ($\rho =0$), there exist 
pseudogap and superconducting (here - BKT) phases  which in fact 
correspond to the cases of absent, uncorrelated and correlated pairs, 
respectively.  Despite the simplicity of the model and assumptions 
made above, the strong dependence of the critical temperatures 
$T_{BKT}$ and $T_{\rho}$ on carrier density is evidently shown.  
Pseudogap phase really exists when $\epsilon_F\leq 
100T_{BCS}^{MF}$ (or $(T_{\rho}-T_{BKT})/T_{\rho}\geq 0.1$ (see 
(\ref{22}))). This conclusion agrees 
qualitatively with the existence of unusual spectral 
peculiarities in some underdoped HTSCs (for example, YBCO 
\cite{Mar1,Din1,Loes1}).  There are many speculations 
about the nature of this pseudogap phase ( in particular, the 
spin-gap conception (see \cite{Pines}) is also very 
popular).  Our results give hopes that "the low-density
phenomena" in HTSCs can be (at least, principally) explained in 
the framework of a rather standard boson-exchange approach though
the simplest phonon model considered can hardly be avaliable
for description of real HTSC compounds. 

It must be also noted 
that unlike this model the 4F one predicts the pseudogap phase  
existence at any fermion densities \cite{Gus1}.  It means 
that in the case 2D boson-exchange model with, as shown above, 
"saturated" critical temperatures as functions of doping, the 
definitions of "underdoped" and "overdoped" samples aquire their 
physical sense.  Nevertheless, a point where both ($T_{\rho}(n_f)$ 
and $T_{BKT}(n_f)$) curves cross each other is absent.  The matter is 
that the second temperature appears at finite value of $\rho$ only. 
 In this connection it is worth to mention that real HTSCs 
 are not pure 2D, but quasi-2D systems with very weak (mainly, 
Josephson) tunneling between neighbouring conducting copper-oxide 
layers. In such a case there can exist one more critical temperature 
\cite{Gor1} 
\begin{equation} T_c\simeq\frac{T_{BKT}}{\ln 
(\epsilon_F|\varepsilon_b|/4t^{2}_{\mid\mid})},\label{23} 
\end{equation}
which (at small $n_f$) defines the formation of usual homogeneous 
condensate where not only $\rho$  but also the phase $\theta$ does 
not  depend on coordinates.  In (\ref{23}) $t_{\mid\mid}$ is the 
inter-plane hopping (tunneling) constant, and $\varepsilon_b$ is the 
two-fermion bound state energy \cite{Lok1,Ran1}.  When $T_c < 
T_{BKT}$ all three temperatures have physical meaning, but when $T_c 
> T_{BKT}$ (or, as it follows from (\ref{23}), 
$t_{\mid\mid}>2\Delta_{BCS}$) the BKT phase cannot be formed and 
pseudogap phase at $T=T_c$ transfers directly to the ordinary 
superconducting phase.  

It cannot be excluded that just such a 
scenario takes place in HTSC copper oxides. Besides, 
superconducting phase which appears at $T=T_{c}$ does not need a 
finite $\rho$, so the curves $T_{\rho}(n_f)$ and $T_c(n_f)$ can in 
principle cross each other in some point $n_f=n_f^{opt}$ which, thus, 
separates two different doping regions. One more result following 
from such a consideration is that the ratio $2\Delta /T_c$ proves to 
be decreasing function of $n_f$  approaching this ratio canonical
BCS value 3.52 at large $n_f$ only what agrees with experimental 
observation (see \cite{Maks}).

Although above it was pointed out that the pseudogap phase -- BKT 
phase transition takes place when $\rho\not= 0$  and fluctuations of 
$\rho$ are not so important as those of $\theta$, it would be 
interesting to estimate their contributions separately.  The 
one-particle gapless spectrum of the normal phase and especially 
pseudogap in abnormal normal phase spectrum are also of great 
interest and need further investigation.

\section*{Acknowledgments}

We would like to thank Prof. V.P.~Gusynin, Dr. E.V.~Gorbar and
Dr. I.A.~Shovkovy for many useful discussions and thoughtful  
comments.

\newpage

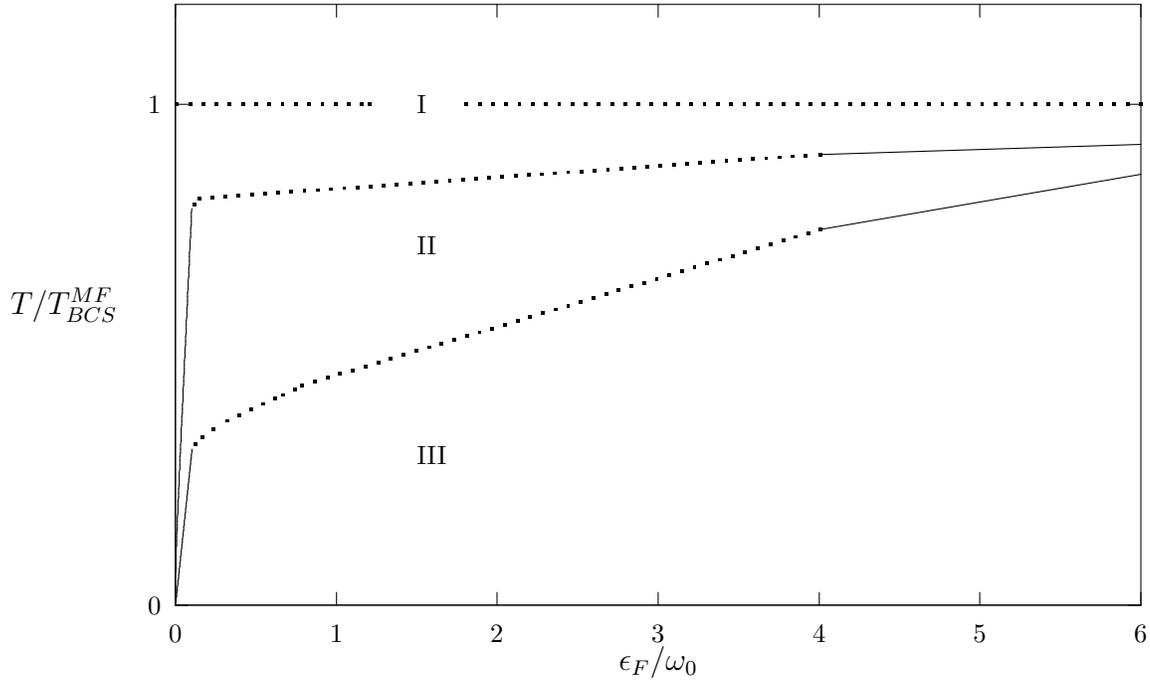
\begin{figure}
\setlength{\unitlength}{0.240900pt}
\ifx\plotpoint\undefined\newsavebox{\plotpoint}\fi
\sbox{\plotpoint}{\rule[-0.200pt]{0.400pt}{0.400pt}}%
\special{em:linewidth 0.4pt}%
\begin{picture}(1800,1080)(0,0)
\font\gnuplot=cmr10 at 10pt
\gnuplot
\put(220,113){\special{em:moveto}}
\put(1736,113){\special{em:lineto}}
\put(220,113){\special{em:moveto}}
\put(220,1057){\special{em:lineto}}
\put(220,113){\special{em:moveto}}
\put(240,113){\special{em:lineto}}
\put(1736,113){\special{em:moveto}}
\put(1716,113){\special{em:lineto}}
\put(198,113){\makebox(0,0)[r]{0}}
\put(220,900){\special{em:moveto}}
\put(240,900){\special{em:lineto}}
\put(1736,900){\special{em:moveto}}
\put(1716,900){\special{em:lineto}}
\put(198,900){\makebox(0,0)[r]{1}}
\put(220,113){\special{em:moveto}}
\put(220,133){\special{em:lineto}}
\put(220,1057){\special{em:moveto}}
\put(220,1037){\special{em:lineto}}
\put(220,68){\makebox(0,0){0}}
\put(473,113){\special{em:moveto}}
\put(473,133){\special{em:lineto}}
\put(473,1057){\special{em:moveto}}
\put(473,1037){\special{em:lineto}}
\put(473,68){\makebox(0,0){1}}
\put(725,113){\special{em:moveto}}
\put(725,133){\special{em:lineto}}
\put(725,1057){\special{em:moveto}}
\put(725,1037){\special{em:lineto}}
\put(725,68){\makebox(0,0){2}}
\put(978,113){\special{em:moveto}}
\put(978,133){\special{em:lineto}}
\put(978,1057){\special{em:moveto}}
\put(978,1037){\special{em:lineto}}
\put(978,68){\makebox(0,0){3}}
\put(1231,113){\special{em:moveto}}
\put(1231,133){\special{em:lineto}}
\put(1231,1057){\special{em:moveto}}
\put(1231,1037){\special{em:lineto}}
\put(1231,68){\makebox(0,0){4}}
\put(1483,113){\special{em:moveto}}
\put(1483,133){\special{em:lineto}}
\put(1483,1057){\special{em:moveto}}
\put(1483,1037){\special{em:lineto}}
\put(1483,68){\makebox(0,0){5}}
\put(1736,113){\special{em:moveto}}
\put(1736,133){\special{em:lineto}}
\put(1736,1057){\special{em:moveto}}
\put(1736,1037){\special{em:lineto}}
\put(1736,68){\makebox(0,0){6}}
\put(220,113){\special{em:moveto}}
\put(1736,113){\special{em:lineto}}
\put(1736,1057){\special{em:lineto}}
\put(220,1057){\special{em:lineto}}
\put(220,113){\special{em:lineto}}
\put(45,585){\makebox(0,0){$T/T_{BCS}^{MF}$}}
\put(978,23){\makebox(0,0){$\epsilon_{F}/ \omega_0$}}
\put(599,900){\makebox(0,0)[l]{I}}
\put(599,679){\makebox(0,0)[l]{II}}
\put(599,349){\makebox(0,0)[l]{III}}
\put(220,113){\special{em:moveto}}
\put(220,119){\special{em:lineto}}
\put(220,126){\special{em:lineto}}
\put(220,132){\special{em:lineto}}
\put(220,138){\special{em:lineto}}
\put(220,144){\special{em:lineto}}
\put(220,151){\special{em:lineto}}
\put(220,157){\special{em:lineto}}
\put(220,163){\special{em:lineto}}
\put(221,170){\special{em:lineto}}
\put(221,176){\special{em:lineto}}
\put(221,182){\special{em:lineto}}
\put(221,189){\special{em:lineto}}
\put(221,195){\special{em:lineto}}
\put(221,201){\special{em:lineto}}
\put(222,207){\special{em:lineto}}
\put(222,214){\special{em:lineto}}
\put(222,220){\special{em:lineto}}
\put(222,226){\special{em:lineto}}
\put(223,233){\special{em:lineto}}
\put(223,239){\special{em:lineto}}
\put(223,245){\special{em:lineto}}
\put(223,251){\special{em:lineto}}
\put(224,258){\special{em:lineto}}
\put(224,264){\special{em:lineto}}
\put(224,270){\special{em:lineto}}
\put(224,277){\special{em:lineto}}
\put(225,283){\special{em:lineto}}
\put(225,289){\special{em:lineto}}
\put(225,296){\special{em:lineto}}
\put(226,302){\special{em:lineto}}
\put(226,308){\special{em:lineto}}
\put(226,314){\special{em:lineto}}
\put(226,321){\special{em:lineto}}
\put(227,327){\special{em:lineto}}
\put(227,333){\special{em:lineto}}
\put(227,340){\special{em:lineto}}
\put(227,346){\special{em:lineto}}
\put(228,352){\special{em:lineto}}
\put(228,358){\special{em:lineto}}
\put(228,365){\special{em:lineto}}
\put(229,371){\special{em:lineto}}
\put(229,377){\special{em:lineto}}
\put(229,384){\special{em:lineto}}
\put(229,390){\special{em:lineto}}
\put(230,396){\special{em:lineto}}
\put(230,402){\special{em:lineto}}
\put(230,409){\special{em:lineto}}
\put(231,415){\special{em:lineto}}
\put(231,421){\special{em:lineto}}
\put(231,428){\special{em:lineto}}
\put(232,434){\special{em:lineto}}
\put(232,440){\special{em:lineto}}
\put(232,447){\special{em:lineto}}
\put(232,453){\special{em:lineto}}
\put(233,459){\special{em:lineto}}
\put(233,465){\special{em:lineto}}
\put(233,472){\special{em:lineto}}
\put(234,478){\special{em:lineto}}
\put(234,484){\special{em:lineto}}
\put(234,491){\special{em:lineto}}
\put(234,497){\special{em:lineto}}
\put(235,503){\special{em:lineto}}
\put(235,509){\special{em:lineto}}
\put(235,516){\special{em:lineto}}
\put(236,522){\special{em:lineto}}
\put(236,528){\special{em:lineto}}
\put(236,535){\special{em:lineto}}
\put(237,541){\special{em:lineto}}
\put(237,547){\special{em:lineto}}
\put(237,554){\special{em:lineto}}
\put(237,560){\special{em:lineto}}
\put(238,566){\special{em:lineto}}
\put(238,572){\special{em:lineto}}
\put(238,579){\special{em:lineto}}
\put(239,585){\special{em:lineto}}
\put(239,591){\special{em:lineto}}
\put(239,598){\special{em:lineto}}
\put(239,604){\special{em:lineto}}
\put(240,610){\special{em:lineto}}
\put(240,616){\special{em:lineto}}
\put(240,623){\special{em:lineto}}
\put(241,629){\special{em:lineto}}
\put(241,635){\special{em:lineto}}
\put(241,642){\special{em:lineto}}
\put(242,648){\special{em:lineto}}
\put(242,654){\special{em:lineto}}
\put(242,661){\special{em:lineto}}
\put(242,667){\special{em:lineto}}
\put(243,673){\special{em:lineto}}
\put(243,679){\special{em:lineto}}
\put(243,686){\special{em:lineto}}
\put(244,692){\special{em:lineto}}
\put(244,698){\special{em:lineto}}
\put(244,705){\special{em:lineto}}
\put(245,711){\special{em:lineto}}
\put(245,717){\special{em:lineto}}
\put(245,723){\special{em:lineto}}
\put(245,730){\special{em:lineto}}
\put(246,736){\special{em:lineto}}
\put(1231,821){\special{em:moveto}}
\put(1736,837){\special{em:lineto}}
\put(221,118){\special{em:moveto}}
\put(221,122){\special{em:lineto}}
\put(222,127){\special{em:lineto}}
\put(222,132){\special{em:lineto}}
\put(223,137){\special{em:lineto}}
\put(223,141){\special{em:lineto}}
\put(224,146){\special{em:lineto}}
\put(224,151){\special{em:lineto}}
\put(225,155){\special{em:lineto}}
\put(225,160){\special{em:lineto}}
\put(226,165){\special{em:lineto}}
\put(226,170){\special{em:lineto}}
\put(227,174){\special{em:lineto}}
\put(227,179){\special{em:lineto}}
\put(228,184){\special{em:lineto}}
\put(228,189){\special{em:lineto}}
\put(229,193){\special{em:lineto}}
\put(229,198){\special{em:lineto}}
\put(230,203){\special{em:lineto}}
\put(230,207){\special{em:lineto}}
\put(231,212){\special{em:lineto}}
\put(231,217){\special{em:lineto}}
\put(232,222){\special{em:lineto}}
\put(232,226){\special{em:lineto}}
\put(233,231){\special{em:lineto}}
\put(233,236){\special{em:lineto}}
\put(234,240){\special{em:lineto}}
\put(234,245){\special{em:lineto}}
\put(235,250){\special{em:lineto}}
\put(235,255){\special{em:lineto}}
\put(236,259){\special{em:lineto}}
\put(236,264){\special{em:lineto}}
\put(237,269){\special{em:lineto}}
\put(237,273){\special{em:lineto}}
\put(238,278){\special{em:lineto}}
\put(238,283){\special{em:lineto}}
\put(239,288){\special{em:lineto}}
\put(239,292){\special{em:lineto}}
\put(240,297){\special{em:lineto}}
\put(240,302){\special{em:lineto}}
\put(241,307){\special{em:lineto}}
\put(241,311){\special{em:lineto}}
\put(242,316){\special{em:lineto}}
\put(242,321){\special{em:lineto}}
\put(243,325){\special{em:lineto}}
\put(243,330){\special{em:lineto}}
\put(244,335){\special{em:lineto}}
\put(244,340){\special{em:lineto}}
\put(245,344){\special{em:lineto}}
\put(245,349){\special{em:lineto}}
\put(246,354){\special{em:lineto}}
\put(246,358){\special{em:lineto}}
\put(1231,703){\special{em:moveto}}
\put(1736,790){\special{em:lineto}}
\sbox{\plotpoint}{\rule[-0.500pt]{1.000pt}{1.000pt}}%
\special{em:linewidth 1.0pt}%
\put(248,742){\usebox{\plotpoint}}
\put(248.00,742.00){\usebox{\plotpoint}}
\multiput(250,745)(12.453,16.604){0}{\usebox{\plotpoint}}
\multiput(253,749)(11.513,17.270){0}{\usebox{\plotpoint}}
\multiput(255,752)(20.704,1.464){47}{\usebox{\plotpoint}}
\put(1231,821){\usebox{\plotpoint}}
\put(250,366){\usebox{\plotpoint}}
\put(250.00,366.00){\usebox{\plotpoint}}
\multiput(260,377)(16.758,12.246){2}{\usebox{\plotpoint}}
\put(299.77,402.43){\usebox{\plotpoint}}
\put(318.55,411.27){\usebox{\plotpoint}}
\put(337.26,420.25){\usebox{\plotpoint}}
\multiput(356,429)(18.916,8.543){2}{\usebox{\plotpoint}}
\multiput(387,443)(18.564,9.282){2}{\usebox{\plotpoint}}
\multiput(417,458)(19.875,5.982){41}{\usebox{\plotpoint}}
\put(1231,703){\usebox{\plotpoint}}
\put(220,900){\usebox{\plotpoint}}
\multiput(220,900)(20.756,0.000){15}{\usebox{\plotpoint}}
\put(523,900){\usebox{\plotpoint}}
\put(675,900){\usebox{\plotpoint}}
\multiput(675,900)(20.756,0.000){52}{\usebox{\plotpoint}}
\end{picture}
\caption{
Phase $T$ -- $n_f$ diagram of 2D metal with indirect inter-carrier
attraction. The solid lines correspond to the calculated 
(for $\lambda =0.5$) parts of
the functions $T_{\rho}(n_{f})$ and $T_{BKT}(n_{f})$, dashed ones
guide on the eye. I, II and III show the regions of the normal, 
abnormal normal and superconducting phases, respectively.  
}
\end{figure}

\end{document}